\documentclass{aastex631}

\definecolor{forestgreen}{RGB}{34,139,34}
\newcounter{qnumber}

\usepackage{ulem}

\begin{document}

\title{Prospects for Reconstructing the Free-floating Planet Mass Function at the Population Level with the Nancy Grace Roman Space Telescope}

\correspondingauthor{William DeRocco}
\email{derocco@umd.edu}

\author[0000-0003-1827-9399]{William DeRocco}
\affiliation{Maryland Center for Fundamental Physics, University of Maryland, College Park, 4296 Stadium Drive, College Park, MD 20742, USA}
\affiliation{Department of Physics \& Astronomy, The Johns Hopkins University, 3400 N. Charles Street, Baltimore, MD 21218, USA}


\author{Samson A. Johnson}
\affiliation{NASA Jet Propulsion Laboratory, Pasadena, CA, 91109, USA}
\affiliation{Department of Astronomy, The Ohio State University, Columbus, OH 43210, USA}

\author{Farzaneh Zohrabi}
\affiliation{Department of Physics and Astronomy, Louisiana State University, Baton Rouge, LA 70803, USA}

\author{Matthew T. Penny}
\affiliation{Department of Physics and Astronomy, Louisiana State University, Baton Rouge, LA 70803, USA}

\author{Peter McGill}
\affiliation{Space Science Institute, Lawrence Livermore National Laboratory, 7000 East Ave., Livermore, CA 94550, USA}



\begin{abstract}
Free-floating planets comprise one of the most enigmatic populations of exoplanets in the Galaxy. Though ground-based observations point to a large abundance of these worlds, little is known about their origins and demographics. In the coming years, the Nancy Grace Roman Space Telescope's Galactic Bulge Time Domain Survey is expected to detect several hundred free-floating planets, providing the first opportunity to characterize these worlds at the population level. We present a first study of Roman's prospects for reconstructing the mass distribution of free-floating planets through population-level statistical inference. We find that depending on the true underlying mass distribution of free-floating planets in the Galaxy, Roman will be able to improve upon existing estimates of the abundance by orders of magnitudes in the largely unexplored mass range below that of Earth. When applied to Roman's observations, the methodology we present herein will be capable of discriminating between different hypothesized mass distributions at high statistical significance, opening a new window into the origins of these rogue worlds.

\end{abstract}

\keywords{}


\section{Introduction} \label{sec:intro}

The Nancy Grace Roman Space Telescope (Roman), NASA's next flagship mission, is poised to transform our understanding of exoplanet demographics. Its Galactic Bulge Time Domain Survey  (GBTDS) will be the most sensitive microlensing survey ever performed and is expected to detect thousands of planets at orbital periods beyond those accessible by transit and radial velocity techniques \citep{Penny2019}. Roman's observations will fill a critical gap in existing demographic studies, providing a rich catalog of planets on intermediate and wide orbits, opening the possibility of detecting analogues of almost every planet in our Solar System \citep{Penny2019}. However, Roman's sensitivity will extend far beyond even the most distant bound planets and into the domain of fully unbound worlds. These are the free-floating planets (FFPs), or, more colorfully, ``rogue worlds,'' planetary mass objects that drift in isolation through the interstellar void. Due to their negligible emission across the electromagnetic spectrum, their detection poses a significant challenge. At the highest masses $(M \gtrsim M_{\text{Jup}})$, direct imaging has been used to identify a large population of young, unbound planetary mass objects in nearby clusters \citep[e.g.,][see \citet{Caballero2018} for a review]{Oasa1999, Zapatero2000, MiretRoig2022}; at lower masses, however, gravitational microlensing provides the strongest means of detecting these worlds. At present, little is known about this population, however the tentative initial detections we have from ground-based microlensing surveys \citep{Mroz2018,Mroz2019,Mroz2020a,Mroz2020b,Ryu2021,Gould2022,Gould2023} suggest that these worlds may dramatically outpopulate their bound counterparts \citep{Sumi2023}, constituting one of the largest, and least understood, demographics of exoplanets in the Galaxy.

Low-mass free-floating planets are primarily thought to be formed in protoplanetary disks and subsequently ejected by the violent dynamical processes that occur during the early stages of system formation. The relative contribution of these different processes is expected to vary as a function of planetary mass; as such, a reconstruction of the free-floating planet mass function would provide a unique window into planetary formation and the underlying physical processes that drive nascent planetary systems towards the quasi-stable system architectures we observe at late times. 

Though Roman is expected to detect hundreds, if not thousands, of free-floating planets during its five-year mission \citep{Johnson2020}, reconstructing the FFP mass function from these observations poses a significant challenge. Microlensing is subject to inherent degeneracies that preclude making a robust mass measurement of an FFP at an event-by-event level \citep{Han2005}. However, the large number of FFP events that Roman is expected to observe will provide the first opportunity to leverage population-level statistics to overcome these inherent event-level degeneracies, as has been successfully applied in related studies on reconstructing the mass function of microlensing isolated black holes \citep{2021arXiv210713697M,2024ApJ...961..179P}. In this paper, we present the first study of Roman's prospects for reconstructing the free-floating planet mass function using population-level statistics. We show that Roman will be able to reconstruct the low-mass end of the FFP mass function for a range of hypothesized mass functions, and will be able to distinguish between these hypotheses at a high degree of statistical significance, providing a new observational probe of the dynamics of young planetary systems throughout the Galaxy.

The paper is organized as follows. In Section \ref{sec:massfuncs}, we begin by describing the various origin scenarios proposed for free-floating planets and outline the current state of observations. In Section \ref{sec:microlensing}, we provide a brief refresher on gravitational microlensing, highlighting the degeneracies that arise when applied to FFPs. In Section \ref{sec:methods}, we discuss our statistical methodology before applying it to simulated Roman data in Sec. \ref{sec:disc} and presenting our results. We conclude with a discussion of future directions in Sec. \ref{sec:conc}.

\section{Free-floating Planets} \label{sec:massfuncs}

Free-floating planets constitute a vast, largely unexplored planetary demographic. With current observations suggesting that Earth-mass FFPs may outnumber their bound counterparts by more than twenty-to-one \citep{Sumi2023}, FFPs are expected to be a ubiquitous outcome of planet formation, possibly constituting one of the largest planetary demographics in the Galaxy. Their origins, however, are unknown, with myriad mechanisms proposed for their formation that depend upon an FFP's mass, age, and location in the Galaxy. At the high-mass end ($M \gtrsim M_{\text{Jup}}$), FFPs have been suggested to form \textit{in situ} in ``star-like'' processes such as direct collapse or aborted gas accretion onto a stellar core \citep{MiretRoig2021}. These FFPs can be difficult to distinguish from the poorly-measured low-mass tail of the stellar mass function \citep{defurio_arxiv}, however recent surveys have pointed to a potential gap in the mass function between the two populations \citep{Gould2022}. At lower masses, FFPs are generally thought to be formed in protoplanetary disks and subsequently ejected by gravitational scattering, for example by a bound planet \citep{Chambers1996,Veras2012,Ma2016,Barclay2017}, an inner binary star system \citep{Smullen2016,Chen2024,Coleman2024}, or a nearby star during a stellar fly-by \citep{Zheng2015,Cai2017,VanElteren2019}. Such ejections may occur early ($\lesssim$ 10 Myr), while the disk is still present and migration is efficient \citep[e.g.][]{Coleman2024}, or much later through long-timescale ($\sim$ Gyr) dynamical instabilities \citep[e.g.][]{Bhaskar2025}. Each of these ejection mechanisms is expected to imprint different features on the mass and velocity distribution of FFPs.

Characterizing FFPs at a demographic level would provide the opportunity to probe a wide range of different astrophysical processes, including low-mass star formation, planetary migration, and the gas properties of protoplanetary disks. Specifically, measuring the mass distribution of these worlds would provide insight into their demographics, and in turn, to the underlying astrophysical processes that birth them. At present, however, this mass function is poorly constrained, due primarily to the small number of FFPs that have been detected by ground-based observatories. Though attempts have been made to fit these observations \citep{Gould2022}, they have largely been restricted to simple functional forms, and even under these restrictions, are highly uncertain. The most up-to-date analysis of existing observations, performed by the Microlensing Observations in Astrophysics collaboration (MOA), fixes the functional form to be a power law and fits for the slope and normalization, finding \citep{Sumi2023}
\begin{equation}
\label{eq:sumi}
    \frac{dN}{d\log M}=2.18^{+0.52}_{-1.40}\times\left(\frac{M}{8M_\oplus}\right)^{-0.96^{+0.47}_{-0.27}}\rm~dex^{-1}~star^{-1},
\end{equation}
This mass function and its associated confidence band\footnote{The plotted confidence interval corresponds to the broken power-law model shown in \citep{Sumi2023}. Note that the confidence band has been shifted downwards by a multiplicative factor in comparison to what appears in the literature due to an error in the version of this plot that appears in \citet{Sumi2023}, confirmed by the authors of \citet{Sumi2023}. The best-fit values are unchanged.} are shown in green in Fig. \ref{fig:moneyplot}, and is uncertain by more than two orders of magnitude in the Earth-mass range. Despite its large uncertainty and fixed, likely unrealistic functional form, we adopt it as a fiducial ``observationally-motivated'' mass function with which to test our reconstruction framework in later sections.

On the theoretical side, there has been a wide range of studies exploring different formation mechanisms, however few of these studies have attempted to use their results to make an explicit prediction for the overall mass function of FFPs in the Galaxy. Notable exceptions include \citet{Ma2016} and \citet{Coleman_2025}, the latter of which found that the theoretically-motivated mass function differs significantly from the power-law mass function fit by MOA at masses just below MOA's present sensitivity. They predict a strong non-monotonic feature at $\approx 8\, M_\oplus$ corresponding to the onset of efficient migration in circumbinary systems; detecting such a feature would provide immediate insight into the extent to which circumbinary systems contribute to the formation of free-floating planets. In the interest of exploring Roman's ability to detect such features, we adopt the \citet{Coleman_2025} mass function as our fiducial ``theoretically-motivated'' mass function in our statistical analysis, recognizing that it is but one possibility for the true underlying mass function.

Other recent works have begun to make additional quantitative predictions for the mass function and to connect them to planet formation theory \citep{Chachan2024, Guo2025, Schib2025}. As these studies mature, they will provide an even richer set of potential features that could be detected in the FFP mass function by Roman. However, as will be discussed in the following section, the inherent degeneracies associated with microlensing signals make reconstructing this mass function a significant challenge. The goal of this paper is to show that such a reconstruction remains achievable and that it will be able to distinguish between various predicted mass functions at a high level of statistical significance, enabling new tests of planet formation theory in settings that are otherwise challenging to explore.

\section{Microlensing} \label{sec:microlensing}

Gravitational microlensing occurs when the gravitational field of a foreground object (the lens) bends the light emitted by a distant luminous source, causing a transient apparent magnification of the source \citep{paczynski_gravitational_1986}. The characteristic angular size of the lensed area is given by its angular Einstein radius, $\theta_E$, defined by 
\begin{equation}
    \theta_E = \sqrt{\frac{4 G M (1 - D_{L}/D_{S})}{ D_{L} \, c^2}},
\end{equation}
where $M$ is the mass of the lens (in our case, an FFP), and $D_{L}$ and $D_{S}$ are the distance from the observer to the lens and source respectively. The typical timescale associated with an event is given by the Einstein crossing time, 
\begin{equation}
\label{eq:tE}
    t_E = \frac{\theta_E}{\mu_{\text{rel}}},
\end{equation}
where $\mu_{\text{rel}}$ is the lens-source relative proper motion. This timescale can also be rewritten in terms of physical lens properties, namely the lens mass $M$, distance $D_L$, and velocity transverse to the line of sight $v_T$ as
\begin{equation}
    t_E = \sqrt{\frac{4 G M D_L (1 - D_{L}/D_{S})}{ v_T^2 \, c^2}}.
\end{equation}
This form for $t_E$ makes clear that there is a family of continuous degeneracies between the underlying physical parameters, hence a measurement of $t_E$ alone is insufficient to measure these parameters individually \citep{Han2005}.

These degeneracies can be partially resolved by the measurement of other features in the light curve. Of particular relevance for FFPs is the finite source parameter $\rho \equiv \theta_S/\theta_E$. Here, $\theta_S = R_S/D_S$, the angular radius of a source with physical radius $R_S$ at a distance $D_S$ from the observer \citep{1994ApJ...430..505W}. For the majority of high-mass FFP events ($M \gtrsim 10-100 \, M_\oplus$), the finite-source parameter is small, hence the light curve is well-approximated by the standard analytic solution for a point-like source and point-like lens. However, for the low-mass FFPs ($M \lesssim \, 10 M_\oplus$) expected to be detected during the GBTDS, the $\theta_S$ and $\theta_E$ begin to become comparable \citep{Johnson2020}. This results in the appearance of finite-source effects in the light curve, namely that the peak magnitude is reduced while the duration of the event is lengthened. In light curves where such effects are detectable, a measurement of $\rho$, along with an estimate of $\theta_S$, provides a means to directly estimate $\theta_E$, resolving one of the degeneracies.\footnote{Note that on an event-by-event level, it is not always possible to estimate $\theta_S$, as this usually accomplished through measurements of the source color and magnitude, or less precisely from the source magnitude alone. However, at the population-level, $\rho$ can be estimated from light curve modeling alone and still retains relevant information about the underlying mass distribution, hence we use $\rho$ in our framework rather than $\theta_E$.}

Only one more independent observable is required to fully break the degeneracy, permitting a direct mass measurement of the lens. The final observable, however, is the most challenging. This is the microlensing parallax \citep{Han2004}
\begin{equation}
    \pi_E = \frac{a_\perp (1-D_L/D_S)}{D_L  \theta_E},
\end{equation}
where $a_\perp$ is the transverse distance between two telescopes that observe the event simultaneously. For higher mass lenses with $t_E \gtrsim$ month, it is possible to measure this parameter using the Earth's orbital motion; however, since FFP events are expected to be short in duration ($t_E \sim 1- 10 ~\text{hours})$, two simultaneous observations are required. Though there are many ongoing efforts to provide such joint observations during Roman's GBTDS using telescopes such as Subaru Hyper-Suprime Cam \citep{Suzuki2025}, DECam \citep{Yang2026}, and Euclid \citep{Bachelet2022,Ban2023}, marshaling such joint observational programs is a significant challenge, hence contemporaneous observations for all identified Roman microlensing events cannot be ensured.

As a result, in order to reconstruct the mass function of FFPs, an exploration of population-level studies is needed. In the following section, we present a methodology by which to do this with the distribution of $t_E$ alone, which is the one assured observable that Roman will measure for all detected events. We then extend the methodology to the joint $t_E - \rho$ distribution and quantify the associated improvement.

\section{Methodology} \label{sec:methods}

\subsection{Overview}

The main principle behind our methodology is to use the observed distribution of $t_E$ from the GBTDS to infer the underlying mass distribution of free-floating planets that gave rise to this observation. In the broadest sense, we wish to fit a model of the \textit{predicted} $t_E$ distribution for a given mass function to the \textit{observed} $t_E$ distribution in order to estimate the underlying mass function associated with the observed data. In order to do this, we first must construct a mapping between FFP mass functions and predicted Roman $t_E$ distributions. We rely on simulations for this step, generating a large number of microlensing events using {\sc gulls}, a Roman-dedicated simulation framework. Using this dataset, we build a model that takes in an FFP mass function and outputs the $t_E$ distribution that Roman would expect to detect in a typical survey. From there, we perform Bayesian inference using a Monte Carlo Markov Chain ensemble sampler to explore parameter space and search for the mass function that best explains the observed $t_E$ distribution. In the following subsections, each of these steps will be discussed in greater detail. Though we have attempted to keep notation clear, we have provided an additional glossary for all symbols that appear in the following sections in Appendix \ref{app:glossary} for easy reference.

\subsection{Simulations}

\label{sec:sims}

To simulate the GBTDS's population of detected free-floating planets from which we will draw inferences we used the {\sc gulls} microlensing simulation code's free-floating planet and Fisher matrix uncertainty modules~\citep{Penny2013,Penny2017,Penny2019,Johnson2020}. {\sc gulls} simulates free-floating planet microlensing events by drawing a lens and source star from synthetic star catalogs generated by a population synthesis Galactic model, replaces the lens star with a free-floating planet, i.e., only its mass and brightness change, but other properties like the distance and proper motion stay the same. As such, we assume the FFP distribution tracks the stellar density and stellar velocity distributions in the Galaxy. {\sc gulls} then simulates the microlensing event with $t_0$ and $u_0$ drawn from  uniform distributions. The photometry is simulated by generating a synthetic image of the scene containing the microlensing event and its surrounding crowded stellar field by, again, using stars drawn from a synthetic star catalog and added to the image using a numerical point spread function. The amount of blending for the event is determined by computing the fraction of flux belonging to the source star within a fixed $0.33$``$\times 0.33$`` square aperture. Photon noise from the magnified source, background stars, and sky background are combined with a Gaussian 1 mmag systematic noise term in quadrature for each lightcurve data point. FFP events are considered detections if, first, the $\Delta\chi^2$ between the a flat lightcurve fit and the input model lightcurve is $\Delta\chi^2=\chi_{\rm flat}^2-\chi_{\rm true}^2>300$, and, second, if there are at least 6 consecutive measurements that are above the flat lightcurve by at least $3\sigma$.

\begin{figure*}
\includegraphics[width=\textwidth]{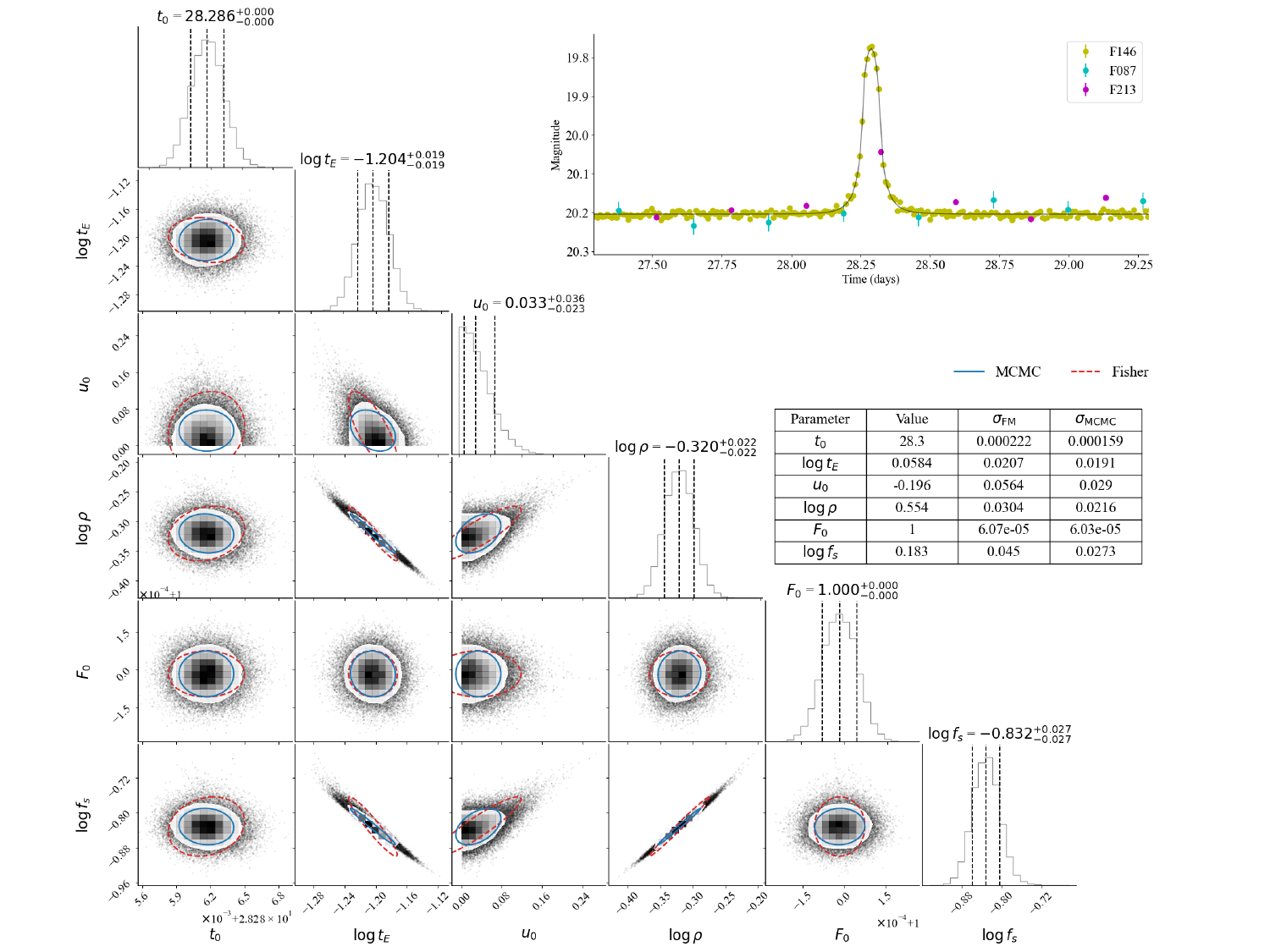}
\caption{Corner plot comparing the free-floating planet lightcurve parameter covariance matrix computed with a Fisher matrix to that computed with Markov Chain Monte Carlo (MCMC) for a simulated free-floating planet detectable by Roman. The parameters are $t_0$ the time of peak magnification, the microlensing timescale $t_{\rm E}$, $u_0$ the impact parameter, $\rho$ the finite source parameter which is the ratio of the source angular diameter to the angular Einstein ring radius, $F_0$ the baseline flux and $f_{\rm s}$, the fraction of the baseline source that belongs to the star being magnified for the F146 filter. The Fisher matrix covariance is represented as dashed red ellipses with major and minor axes determined by the eigenvectors of the mini-covariance matrix of the two plotted parameters extracted from the full covariance matrix. The covariance matrix of a trimmed and thinned MCMC chain is represented as solid blue ellipses. The grayscale map shows the density distribution of the trimmed and thinned chain. To enable better comparison of the covariances, the Fisher ellipses have been shifted to the median of the MCMC chain, but were evaluated at the true value of the parameters, indicated by a red dot. Histograms show the marginalized MCMC distribution for each parameter. The table lists the parameter values and their uncertainties estimated with each method. The inset at the top right shows the simulated lightcurve data for this corner plot, with colored data points representing observations in different filters, and the solid line showing the true lightcurve.}
\label{fig:fisher}
\end{figure*}

Following the same method as \citet{Penny2017}, for every detected event a Fisher matrix analysis is used to estimate the covariance matrix of the event's parameters $t_0$, $u_0$, $t_{\rm E}$, $\rho$, and two flux parameters, the baseline flux $F_{\rm base}$ and the blend fraction $f_{\rm s}=F_{\rm s}/F_{\rm base}$, where $F_{\rm s}$ is the source flux. To construct the Fisher information matrix we compute analytic derivatives of the \citet{witt_can_1994} finite source lightcurve with respect to each parameter. These are weighted by the inverse of the photometric uncertainties and the product for pairs of parameters summed for each element of the matrix~\citep[e.g.,][]{Gould2003}. The Fisher information matrix is inverted to obtain the covariance matrix. The uncertainty on each parameter is then the square root of the parameter's term on the diagonal of the covariance matrix. Note that this uncertainty already includes the effects of degeneracies in fit parameters, e.g. degeneracies between blend fraction and the impact parameter \citep{Distefano1995,Wozniak1997,Alcock1998}.

As the Fisher matrix method assumes Gaussian uncertainties and a parabolic local minimum of $\chi^2$, in Figure~\ref{fig:fisher} we use Markov Chain Monte Carlo analysis to validate its uncertainty estimates for a simulated event with strong finite source effects. 
The Fisher- and MCMC-computed uncertainties and their covariances are in broad agreement. 
Note that the Fisher matrix overestimates the uncertainties on the parameters that are degenerate with $u_0$ ($\rho$, $t_{\rm E}$, and $f_{\rm s}$) because we applied a $u_0 > 0$ prior on the MCMC evaluation that leads to an asymmetric non-Gaussian posterior. Had we allowed negative $u_0$, the width of the MCMC marginalized posterior of the degenerate parameters would not be significantly changed, because the spread into negative $u_0$ solutions would just be reflected over the axis to form a banana-shaped posterior of comparable width to the asymmetric posterior seen in our analysis. 
Note that this feature of the posterior only appears in events with strong finite source effects in which $|u_0|$ is consistent with zero
\citep[e.g.,][]{Johnson_2020}. 
The agreement between Fisher matrix and MCMC uncertainties is better for events without finite source effects because they do not typically have posteriors shaped by the priors. 
However, in all cases, priors on parameters (such as the blending parameter $f_s$ and impact parameter $u_0$) in the MCMC will tend to reduce the uncertainty in parameters relative to the prediction of the Fisher matrix, so our use of the Fisher matrix errors to estimate the parameter uncertainty is 
conservative.

Using the above procedure we generated 990,000 simulated Roman microlensing events, of which $N_{\text{sim}} = 906,749$ are FFPs. Each event is weighted by a $w_i\propto2\mu_{\rm rel}\theta_{\rm E}$ that defines the relative contribution of a particular event to the overall expected event rate in Roman. The $w_i$ are normalized such that $\sum_{i=1}^{N_{\text{sim}}} w_i = \Gamma_{\text{GBTDS}}$, where $\Gamma_{\text{GBTDS}}$ is the predicted FFP event rate during Galactic Bulge Time Domain Survey. The lenses are drawn from a full Galactic population of FFPs, brown dwarfs, stars, and stellar remnant provided by SynthPop \citep{SynthPop}. This model utilizes a spatio-kinematic disk model developed by \citep{koshimoto_parametric_2021}, the bulge density distribution of \citep{Cao2013}, and the extinction map of \citet{Surot2020}, and shows agreement to within 20\% of \citeauthor{Mroz2019b}'s (\citeyear{Mroz2019b}) OGLE microlensing event rates~\citep{Huston2026}. The brown dwarfs are initially sampled from a log-uniform distribution in mass, then subsequently reweighted so that their mass function matches the \citet{Kroupa2001} initial mass function (IMF), $dN/dM\propto M^{-0.3}$ in the range $0.013\le M/M_{\odot} < 0.08$ so that it is continuous with the \citet{Kroupa2001} IMF for stars. 

While the stars are drawn proportionally to their modeled Galactic abundance, the free-floating planets are instead sampled from a log-uniform distribution in mass. Though this mass function is not observationally nor theoretically motivated in and of itself, it provides a simple baseline by which to rescale the event weights for any mass function of interest, i.e. $\tilde{w}_i = w_i\Phi(M_i)$, where $w_i$ is the {\sc gulls} weight for a log-uniform distribution and $\tilde{w}_i$ is the event weight multiplied by the mass function defined by $\Phi(M) \equiv \frac{dN_{\text{FFP}}}{d\log_{10}M}(M)$.

\begin{figure}[ht!]
\plotone{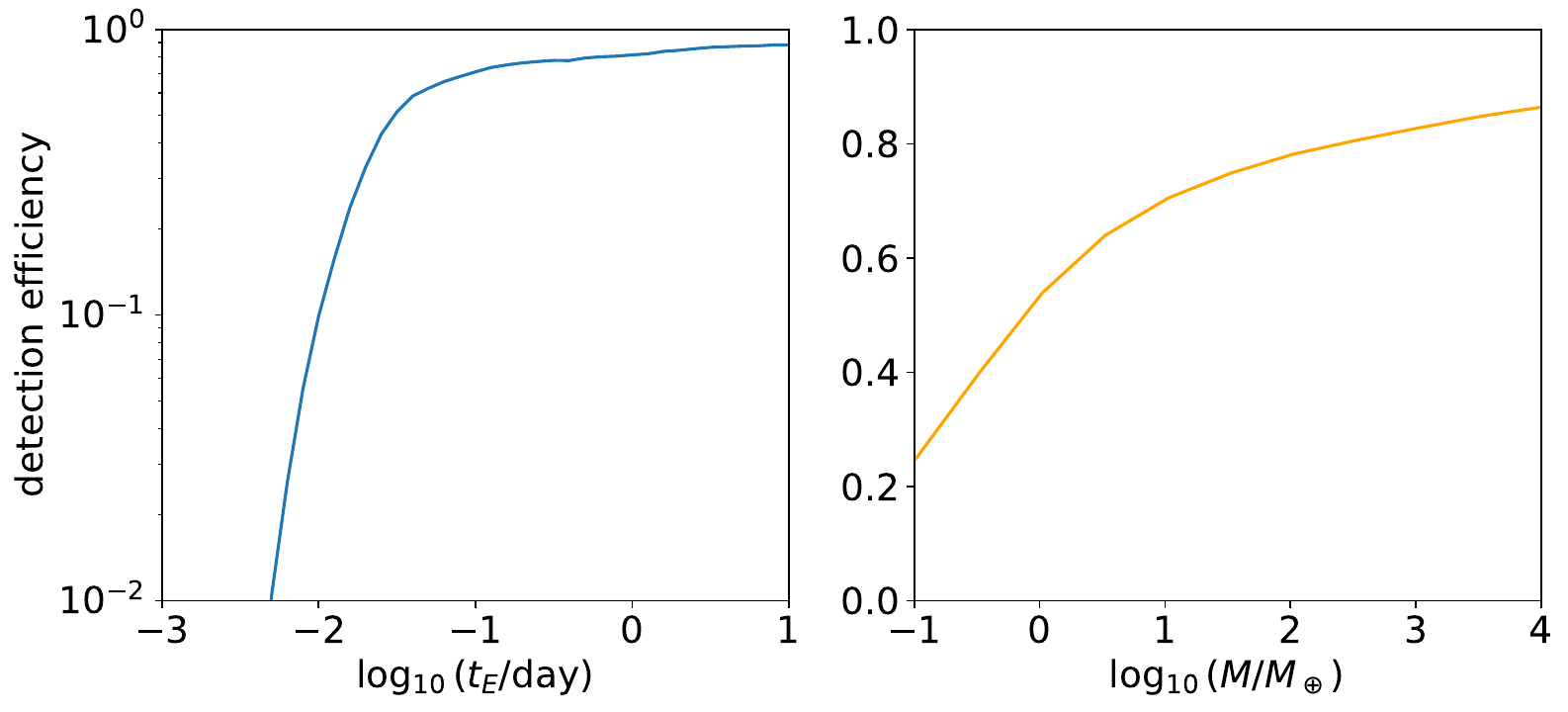}
\caption{Detection efficiency as a function of both $t_E$ (left) and lens mass (right). The $t_E$ plot assumes a log-uniform mass function for FFPs. Our results agree with previous estimates of Roman's FFP detection efficiency presented in \citet{Johnson_2020}.}
\label{fig:deteff}
\end{figure}

The {\sc gulls} simulation therefore provides an appropriately weighted sample of FFP microlensing events that Roman is expected to observe. Furthermore, {\sc gulls} generates light curves for each of these events given Roman's current performance benchmarks, providing useful quantities such as $N_{3\sigma}$, the number of observations that are at least $3\sigma$ above baseline variation in the light curve, and $\Delta \chi^2$, which measures the difference in goodness-of-fit between a flat model for the light curve and a point-source point-lens microlensing curve \citep{paczynski_gravitational_1986}. These quantities allow us to place detection cuts in keeping with expectations for Roman's detection pipeline to isolate a sample of detectable events in the {\sc gulls} simulation output. We place our detection threshold at $N_{3\sigma} \geq 6$ and $\Delta \chi^2 > 300$ \citep{Johnson2020}, which results in a sub-sample of $N^{\text{det}}_{\text{sim}} = 430,252$ simulated events from the initial 906,749.\footnote{While real data may include additional cuts, e.g. cuts associated with data quality, the consequence of such cuts is mainly to reduce the sample of stars available to search,  not the relative detection efficiency as a function of $t_E$ \cite{Koshimoto2023}. As such, the inclusion of such cuts would not be expected to have a large impact on the overall shape of the recovered mass function.} Again, it is important to keep in mind that these simulated ``events'' are just pairs of lenses and source stars, and do not correspond to the actual distribution of predicted events. The true, predicted number of detected events as seen in six seasons of Roman observations is found by taking the 430,252 and summing their weights, i.e. $\sum^{N^{\text{det}}_{\text{sim}}}_{i=1} w_i \Phi(M_i) = N^{\text{det}}_{\text{pred}}$, then multiplying by the survey baseline, which we take to be six 72-day seasons of continuous observation at 15-minute cadence \citep{Penny2019}. The associated detection efficiency as a function of $t_E$ can then be easily computed as $\sum^{N^{\text{det}}_{\text{sim}}}_{i=1} w_i[{\rm (detection~criteria)}~\&~(t_E~{\rm bin~limits})]/\sum^{N^{\text{det}}_{\text{sim}}}_{i=1} w_i$, where the square brackets represent a selection operation. The detection efficiency is shown as a function of both $t_E$ (left) and lens mass (right) in Fig.~\ref{fig:deteff}. We see that the majority of events that do not pass selection cuts are low-mass events with $t_E$ too short to provide 6 points above $3\sigma$ in the light curve.

We display the resulting expected yields of detected FFPs per dex in Table \ref{tab:yields} for three fiducial mass functions. Note that the log-uniform yields differ from those presented in \citet{Johnson2020} due to updates to the field placement, cadence, and Galactic model.

\begin{table}
    \centering
    \begin{tabular}{l|cccccc|c}
        Mass Function & $<$ & $0.1\, M_{\oplus}$  & $1\, M_{\oplus}$ & $10\, M_{\oplus}$ & $100\, M_{\oplus}$ & $1000\, M_{\oplus}$ & Total detected events\\
        \hline\hline
        MOA & 18 & 253 & 330 & 163 & 62 & 112 & 1010\\
        Coleman & 0 & 1 & 6 & 41 & 22 & 93 & 233\\
        Log-uniform & 0 & 2 & 15 & 64 & 219 & 757 & 1057 \\
    \end{tabular}
    \caption{The expected number of detected FFP events for the full Roman Galactic Bulge Time Domain Survey given our fiducial mass functions. The first row corresponds to the best-fit power law described in \citet{Sumi2023}, the second row to the simulation-based mass function described in \citet{Coleman_2025}, and the third row to a log-uniform distribution of 1 FFP per star per dex. Here, the middle columns count the expected number of events in the dex-width bin centered at the specified value, with the $<$ to the underflow bin. These values are taken from simulations that assume the Galactic model presented in \citet{SynthPop}.}
    \label{tab:yields}
\end{table}

\subsection{$t_E$ distribution model}

With the microlensing timescale typically being the only observable quantity containing information about the mass of the lens, the usual practice is to infer a mass function from a distribution of microlensing timescales by forward modeling the timescale distribution for a model mass function with free parameters and an assumption of a Galactic density and kinematic model for the lenses and sources. After applying the detection cuts of the previous subsection, we have a set of events for which both the observable parameters and underlying parameters are known. As such, it is simple to construct a ``truth'' $t_E$ distribution for any choice of FFP mass function $\Phi(M)$ by applying the mass function to the weights as described above, then constructing a $t_E$ histogram of the events using these modified weights. 

Past studies \citep[e.g.,][]{Sumi2011,sumi2023freefloating} have modeled the FFP mass function as a power law, but here we wish to be more general, and instead define it as the smooth interpolation through a finite number of points spaced equally in logarithmic mass. Let us define $\{n^{t_E}_j\}[\Phi(M)]$ as the expectation value for the number of FFP events that Roman will detect in a $t_E$ bin indexed by $j$, where $j$ runs from 1 to $N^{t_E}_{\text{bins}}$. This set of $\{n^{t_E}_j\}[\Phi(M)]$ defines our model, and is clearly dependent on the choice of $\Phi(M)$. Therefore, given an observed distribution $\{n^{t_E}_j\}^{\text{obs}}$, we seek to select a mass function $\Phi(M)$ such that $\{n^{t_E}_j\}[\Phi(M)]$ best fits the observed distribution $\{n^{t_E}_j\}^{\text{obs}}$. However, $\Phi(M)$ is a continuous function, the form of which we wish to remain as agnostic as possible about. Therefore, we choose to parameterize $\Phi(M)$ by selecting values of this function at a particular set of $N^{\text{mass}}$ masses that are logarithmically spaced and interpolating between these values, hence $\Phi(M)$ is in practice modeled by $N_{\text{mass}}$ parameters, where $N_{\text{mass}}$ is the number of mass bins. The model can therefore be written as $\{n^{t_E}_j\}(\{\Phi(M_k)\})$ where $k$ runs from 1 to $N^{\text{mass}}$.

There are clearly a variety of hyperparameters associated with this model, most notably the choice of the number of bins in $t_E$ histogram ($N^{t_E}_{\text{bins}}$) and the number of masses $M_k$ with which to parameterize the mass function ($N^{\text{mass}}$). There are two limits for each of these choices. As $N^{t_E}_{\text{bins}} \rightarrow \infty$, the number of expected events in any given bin drops well below one, resulting in large sampling variation and a corresponding increase in the relative uncertainty; as $N^{t_E}_{\text{bins}} \rightarrow 1$, the model loses all ability to discern features in the distribution and can only be used to constrain the overall number of expected events, leading to a strong degeneracy in the various parameters of the mass function. Similarly, as $N^{\text{mass}} \rightarrow \infty$, the mass function can be modeled arbitrarily well, however the relative importance of nearby masses becomes difficult to distinguish, leading to a broad and strongly covariant posterior for nearby masses (see Sec. \ref{sec:binconfusion}); as $N^{\text{mass}} \rightarrow 2$, the reconstructed mass function becomes restricted to being a simple power-law over a finite interval in log-log space. It is clear that there are optimal intermediate values for both of these hyperparameters that depend on the true underlying mass function. Given that with true Roman data, we will not know the underlying mass function, we have instead explored various values of these hyperparameters to develop reasonable heuristics that provide a good balance of mass function resolution and low per-parameter uncertainty (see Sec. \ref{sec:disc}). Our fiducial reconstruction ultimately adopts $N^{\text{mass}} = 8$ values over the range $10^{-1}\, M_\oplus$ to $10^{6}\, M_\oplus$ at dex intervals and $N_{\text{bins}}^{t_E} = 20$, with $t_E$ bins equally spaced on a logarithmic scale between $t_E = 10^{-3}$ and $10^{3}$ days.

This framework is not necessarily optimized to discerning specific features of interest in the FFP mass function, and more flexible frameworks will be explored in future work. However, for certain models, such as the \citet{Sumi2023} power law, we find that this simple framework is already sufficient to reconstruct broad features in the mass function, providing a critical tool for discriminating between different hypotheses on the abundance and origins of free-floating worlds.

\subsection{Model-fitting}

With this simulation-based model in hand, we can write down a likelihood function. We assume Gaussian uncertainties on the counts in each $t_E$ bin corresponding to the Poissonian scale of fluctuations, i.e. $\sigma_j^{0} \equiv \sqrt{n^{t_E, \text{obs}}_j}$. Note that this uncertainty only captures the statistical uncertainty associated with bin fluctuations, not underlying uncertainties in the measurement of an individual $t_E$. We additionally include this uncertainty by treating each measurement $t_{E, i}$ and its associated uncertainty $\delta t_{E,i}$ (taken to be the square root of the $(t_E, t_E)$ entry of the inverse Fisher matrix provided by {\sc gulls}) as a Gaussian-distributed random variable sampled from $\mathcal{N}(t_{E, i}, \delta t_{E,i})$.

The individual uncertainties on each $t_E$ measurement can be used to compute a relative probability $p_{i,j}$ for the event $i$ being measured in any given $t_E$ bin indexed by $j$. The uncertainty on the height of any bin $j$, is given by $\sigma_j^{\delta t_E} \equiv \sqrt{\sum_i p_{i,j} \tilde{w}_i}$, which, in the limit that $\delta t_{E,i} \rightarrow 0 ~\forall i$ just yields the Poisson uncertainty on bin fluctuation alone, $\sigma_j^{0}$. In principle, this procedure should only serve to spread out each event such that the overall expected value in each bin decreases, resulting in a larger relative uncertainty on that bin height. However, due to the necessity of imposing a cutoff on the high mass and low mass end of the FFP mass function when using {\sc gulls}, the prescription of using the expected values actually results in artificially smaller uncertainties in bins near the edges. As such, we choose to adopt a conservative prescription for the per-bin uncertainty by taking the larger of the two relative uncertainties for each bin and multiplying by the observed number of events in the bin. We denote this final, conservative uncertainty $\sigma_j$.

\begin{figure}[ht!]
\plotone{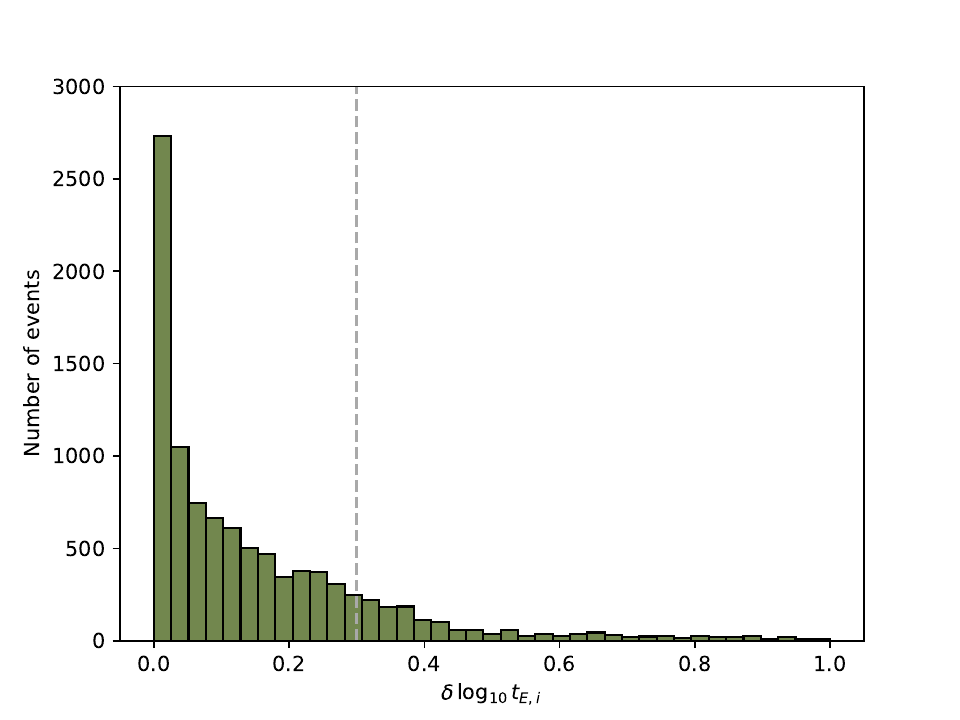}
\caption{Histogram of detected events for a log-uniform mass function showing the uncertainty on individual $t_E$ measurements from modeling, $\delta \log_{10} t_{E,i}$. The gray dashed line corresponds to the fiducial $t_E$ bin width used for reconstructing the mass distribution.
\label{fig:deltatE}}
\end{figure}

Ultimately, the inclusion of these per-event uncertainties has little effect on the mass function reconstruction. This can be understood from Fig. \ref{fig:deltatE}, which shows the distribution of uncertainties on $\log_{10} t_E$ from the model fitting, $\delta \log_{10} t_{E,i}$. A gray dashed line is shown at the fiducial $t_E$ bin width we use in our analysis. It is clear that the majority of events have uncertainties well below the bin width. As such, these fitting uncertainties do not appreciably affect the overall uncertainty on the reconstructed mass function, which is dominated by Poisson uncertainty due to low-number statistics. For ease of comparison, we adopt the pure Poissonian uncertainty in the comparative studies presented in Secs. \ref{sec:binconfusion} -- \ref{sec:syst}.

The likelihood is defined as 
\begin{equation}
\label{eq:likelihood}
    \mathcal{L}(\Phi(M_k)|\{n^{t_E}_j\}^{\text{obs}}) = \prod_j (\sqrt{2\pi} \sigma_j)^{-1} \exp\left[-\frac{1}{2}\left(\frac{n^{t_E, \text{obs}}_j - n^{t_E}_j(\Phi(M_k))}{\sigma_j}\right)^2\right]
\end{equation}
We use \texttt{emcee}, a Monte Carlo Markov Chain ensemble sampler to explore parameter space and provide estimates for the underlying values of $\Phi(M_k)$. We adopt a uniform prior of $\Phi(M_k) \in [0, \infty)$ and run the sampler with 32 walkers for 250,000 steps each with a burn in period of 50,000 steps. We check that these runs converge by ensuring that chain length at the end of the run exceeds the autocorrelation length by a factor of $>100$ and that the estimate for the autocorrelation length has remained within $1\%$ for at least 20,000 steps. The result is a Bayesian posterior over the $N^{\text{mass}}$-dimensional parameter space spanned by the $\Phi(M_k)$. From this posterior, one can easily compute confidence intervals for each value of $\Phi(M_k)$ by marginalizing over the remaining $N^{\text{mass}}-1$ points in the mass function. Applying this procedure, we reconstruct an estimated mass function and associated confidence intervals on its normalization at each mass.

\section{Discussion} \label{sec:disc}

\subsection{Primary results}

\begin{figure}[ht!]
\plotone{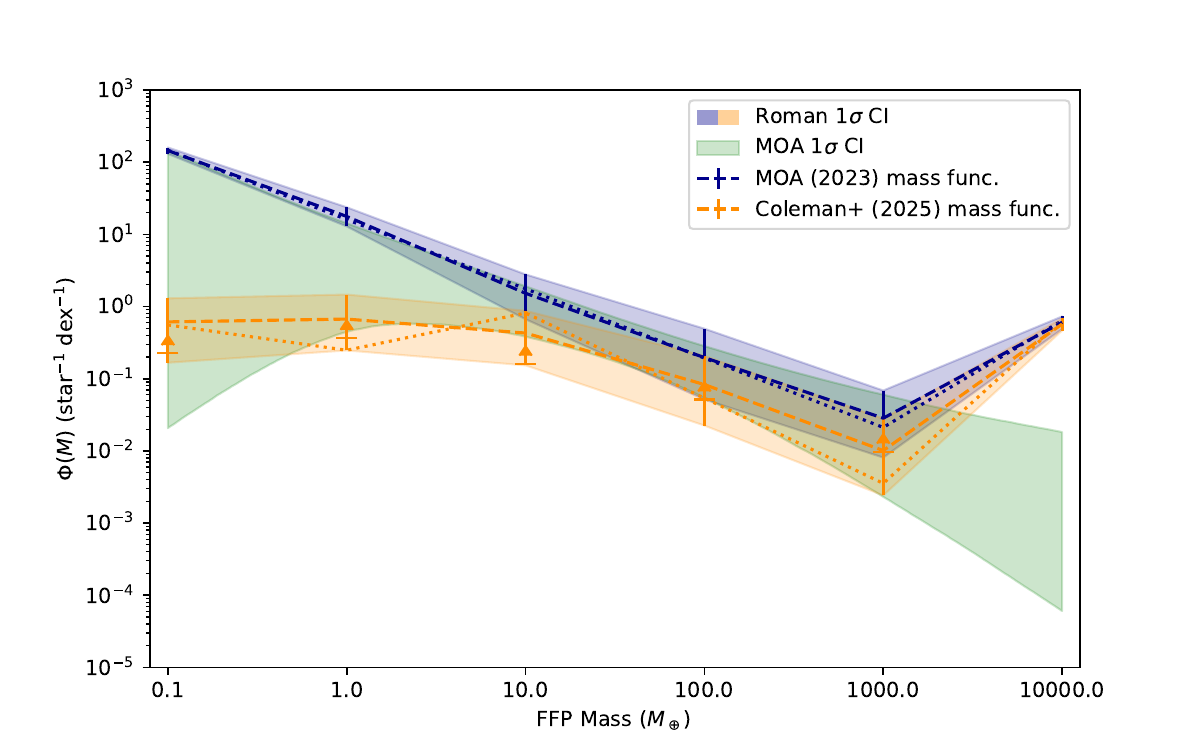}
\caption{Reconstructed free-floating planet mass functions for our fiducial values of $N^{\text{mass}} = 8$ and $N_{\text{bins}}^{t_E} = 20$, where only the lower six mass points are shown. The blue curve corresponds to the mass function presented in \citet{Sumi2023} (see Eq. \ref{eq:sumi}) and the orange to the simulation results of \citet{Coleman_2025}. In both cases, the shaded regions show the $1\sigma$ (68\%) confidence intervals on the reconstructed values. Arrows indicate the threshold below which the reconstruction is dominantly prior-supported, hence the lower limits at those points are statistically indistinguishable from zero. The current $1\sigma$ confidence interval on a broken power law mass function from MOA observations \citep{Sumi2023} is shown in green for comparison. 
\label{fig:moneyplot}}
\end{figure}

Our primary results are shown in Fig. \ref{fig:moneyplot}. This plot displays the $1\sigma$ (68\%) confidence interval (shaded regions) of the reconstructed FFP mass function for two fiducial models: the observationally-motivated power-law fit presented in \citet{Sumi2023} (blue dashed) and the planet population synthesis simulation-based theoretical mass function predicted in \citet{Coleman_2025} (orange dashed). (For further discussion of these two mass functions, see Sec. \ref{sec:massfuncs}.) The current $1\sigma$ confidence limit as presented in \citet{Sumi2023} is shown as a green band, though recall that the fit performed in \citet{Sumi2023} used a more restricted functional form and adopted different priors to our analysis, hence cannot be directly compared to our results. Additionally, we include upward-pointing arrows on our reconstruction indicating the region in which the results are likelihood-supported rather than prior-supported. (We suppress these arrows for masses at which the $1\sigma$ confidence interval on the reconstruction is fully likelihood-supported.) Mass values on which these arrows appear are statistically indistinguishable from zero at the level of our analysis.\footnote{Though our prior is formally semi-infinite for each value of $\Phi(M_k)$, we choose to estimate the region of prior support as the lower limit on the confidence region of a uniform prior between zero and twice the maximum sample returned for a given $\Phi(M_k)$.} We have chosen to adopt $N^{\text{mass}} =8$ and $N_{\text{bins}}^{t_E} = 20$ as our fiducial hyperparameter values, however we show only the lowest 6 mass points, corresponding to the FFP and brown dwarf mass range, in the plots below. We find that these values provide a good balance of mass resolution and low statistical uncertainty per selected mass for the mass functions we study. 
The reconstructed $\Phi(M_k)$ for the FFP mass range are tabulated numerically in Table \ref{tab:massfuncs}.

\begin{table}
    \centering
    \begin{tabular}{l|ccccc}
       Mass Function & $\Phi(0.1\, M_{\oplus})$  & $\Phi(1\, M_{\oplus})$ & $\Phi(10\, M_{\oplus})$ & $\Phi(100\, M_{\oplus})$ & $\Phi(1000\, M_{\oplus})$ \\
        \hline\hline
        MOA & $145^{+14}_{-15}$ & $17^{+6}_{-5}$ & $1.5^{+1.3}_{-0.9}$ & $0.20^{+0.3}_{-0.15}$ & $0.03^{+0.04}_{-0.02}$ \\
        ~ & 147 & 16 & 1.8 & 0.2 & 0.02 \\
        Coleman & $0.6^{+0.7}_{-0.5}$ & $0.7^{+0.8}_{-0.4}$ & $0.4^{+0.4}_{-0.3}$ & $0.08^{+0.12}_{-0.06}$ & $0.01^{+0.02}_{-0.01}$ \\
        ~ & 0.6 & 0.3 & 0.8 & 0.05 & 0.004 \\
        Log-uniform & $1.1^{+1.1}_{-0.8}$ & $1.3^{+1.4}_{-0.8}$ & $0.8^{+0.8}_{-0.5}$ & $1.1^{+0.5}_{-0.4}$ & $1.0^{+0.3}_{-0.2}$ \\
        ~ & 1 & 1 & 1 & 1 & 1 \\
    \end{tabular}
    \caption{Tabulated values of the reconstructed mass function $\Phi(M) = dN/d\log_{10}M$ in per star per dex at various masses. The first row corresponds to the best-fit power law described in \citet{Sumi2023}, the second row to the simulation-based mass function described in \citet{Coleman_2025}, and the third row to a log-uniform distribution of 1 FFP per star per dex. Quoted uncertainties corresponds to the $1\sigma$ (68\%) confidence interval on the value. Truth values are shown for ease of comparison beneath each reconstruction row.}
    \label{tab:massfuncs}
\end{table}

The main takeaway from these results is that with $t_E$ alone and a low number of mass bins, Roman's observations will provide the opportunity to statistically discriminate between different potential FFP mass functions, especially at masses well below that of Earth. With models predicting a wide range of possible forms for the mass function, this will enable Roman to test specific hypotheses on the origins and growth of FFPs. At the same time, we find that models that predict very few events, such as that of \citet{Coleman_2025}, will present a significant challenge to reconstruction at the population level due to limited statistics. As such, while the methodology presented in this paper is a useful tool, it is complementary to and provides further motivation for the joint observational campaigns discussed in the Introduction.

In the following subsections, we will discuss various sources of uncertainty that are introduced both by our statistical methodology and aspects of our modeled Galactic populations, and describe each of their impacts on Roman's prospects for reconstructing the FFP mass function.

\subsection{Brown dwarf contamination}

Given that we do not have direct mass measurements of detected microlensing events, it is possible for higher-mass isolated lenses to contaminate FFP detections. In particular, fast-moving or distant/nearby brown dwarfs with masses $M > 13\,M_{\text{Jup}}$ can mimic FFP events due to the inherent degeneracy in $t_E$ between lens mass, distance, and proper motion (see Eq. \ref{eq:tE}). The inclusion of brown dwarfs in our Galactic population, as described in Sec. \ref{sec:sims}, leads to a considerable increase in the uncertainty on our reconstructed FFP mass function, in particular at the highest masses. While the uncertainties only increase with the inclusion of brown dwarfs by a factor of $\sim2-3$ for masses below $1000\,M_\oplus$, brown dwarf confusion can lead to an increase in the uncertainty on the order of $\sim 10$ at $1000\,M_\oplus$. It is not surprising that the dominant effect of brown dwarf contamination appears at the highest masses, as there are significantly more brown dwarfs that have $t_E$ values that are comparable to those of FFPs in that mass range. 
As can be seen from the quoted uncertainties in Table \ref{tab:massfuncs}, the combined effects of low-number statistics and brown dwarf contamination present a challenge for accurately reconstructing the FFP mass function at $M \gtrsim 10\,M_\oplus$, motivating joint observational campaigns to provide direct measurements of the underlying lens parameters.

\subsection{Bin confusion}
\label{sec:binconfusion}

As discussed in Section \ref{sec:methods}, varying $N^{t_E}_{\text{bins}}$ and $N^{\text{mass}}$ will change the overall uncertainty and mass resolution of the reconstructed mass function. This is particularly relevant in the case of $N^{\text{mass}}$, for which the adoption of a large number of parameterized values can lead to what we term ``bin confusion.'' This occurs when $M_k$ and $M_{k+1}$ are sufficiently close to one another that the effects of varying $\Phi(M_k)$ and $\Phi(M_{k+1})$ are statistically indistinguishable, leading to a strong degeneracy between nearby values. This inflates the associated uncertainty on any given $\Phi(M_k)$.

\begin{figure}[ht!]
\plotone{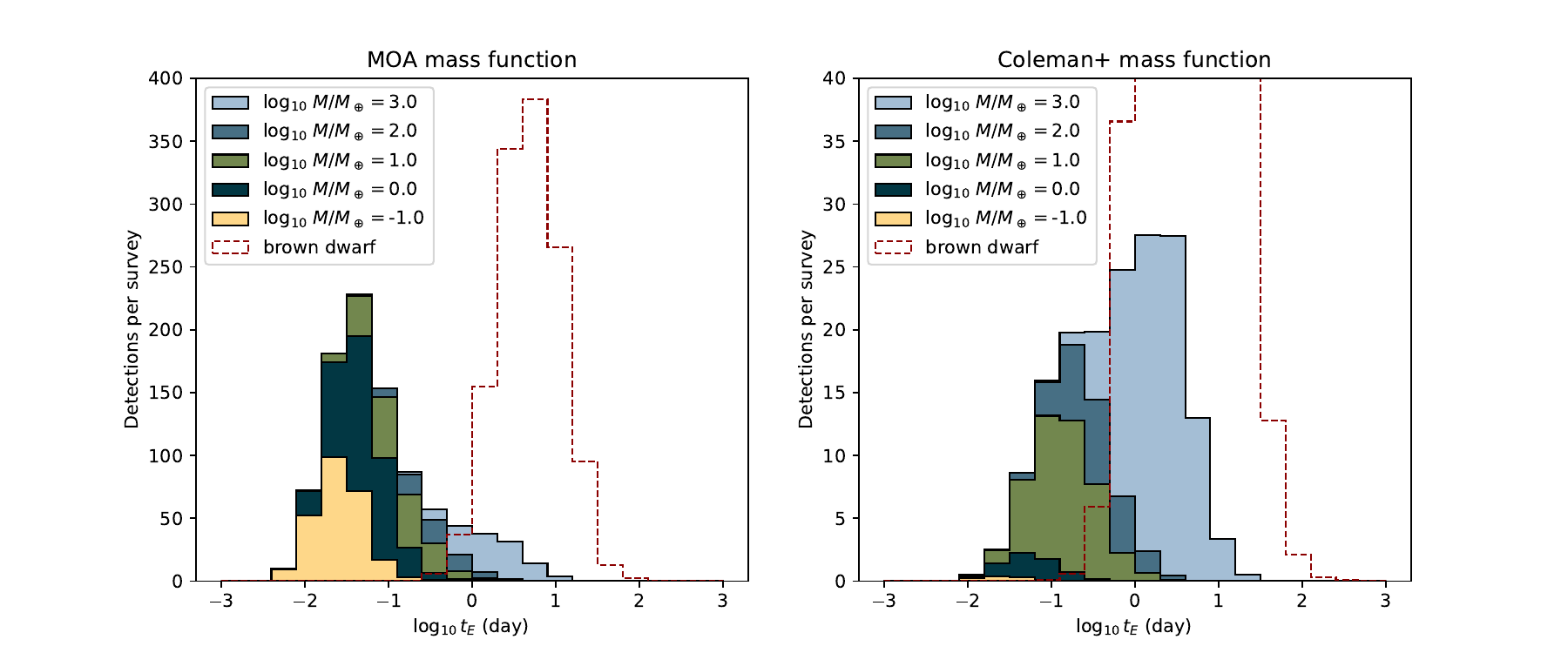}
\caption{Stacked histogram of detected events showing the respective contribution to each $t_E$ bin of FFPs of different masses assuming the MOA FFP mass function (Eq. \ref{eq:sumi}) (left) and \citet{Coleman_2025} mass function (right). Here, $N^{\text{mass}} = 8$ and $N_{\text{bins}}^{t_E} = 20$. The red dashed histogram corresponds to the expected contribution from brown dwarfs, which dominates at $t_E \gtrsim 1$ day.}
\label{fig:tEhists}
\end{figure}

Useful plots to diagnose this behavior are of the form shown in Fig. \ref{fig:tEhists}. These plots show the $t_E$ distribution for the truth mass function as a stacked histogram, where each color contains the events closest in log-space to a particular $M_k$ in mass. At a qualitative level, the way to interpret one of these plots is that the most strongly constrained masses will be the ones that dominate in at least one bin of the $t_E$ distribution. This is because, if one mass $M_k$ dominates in a bin, then variations in $\Phi(M_k)$ will lead to large relative changes in the height of the corresponding $t_E$ bin. With this in mind, we see from Fig. \ref{fig:tEhists} that the most constrained masses will be $0.1\, M_\oplus$ (tan, dominating in the $t_E = 10^{-2}$ day bin) and $10\, M_\oplus$ (green, dominating in the $t_E = 10^{-1.5} - 10^{-0.5}$ day bins) for the the MOA and Coleman mass functions, respectively. Comparing to Fig. \ref{fig:moneyplot}, we see that these values do in fact correspond to the masses with lowest uncertainty in the reconstruction, as expected. We have additionally shown the predicted contribution to the $t_E$ distributions from the brown dwarf population, which dominates the FFP distribution at $t_E \gtrsim 1$ day. This leads to increased uncertainty on the mass reconstruction of the FFPs, and, is largely responsible for the larger uncertainty in the $\sim1000 M_\oplus$ range seen in Fig. \ref{fig:moneyplot}, even though by the logic above, this mass bin should be well-constrained since it dominates the FFP distribution in the $t_E > 1$ day bins.

\begin{figure}[ht!]
\plotone{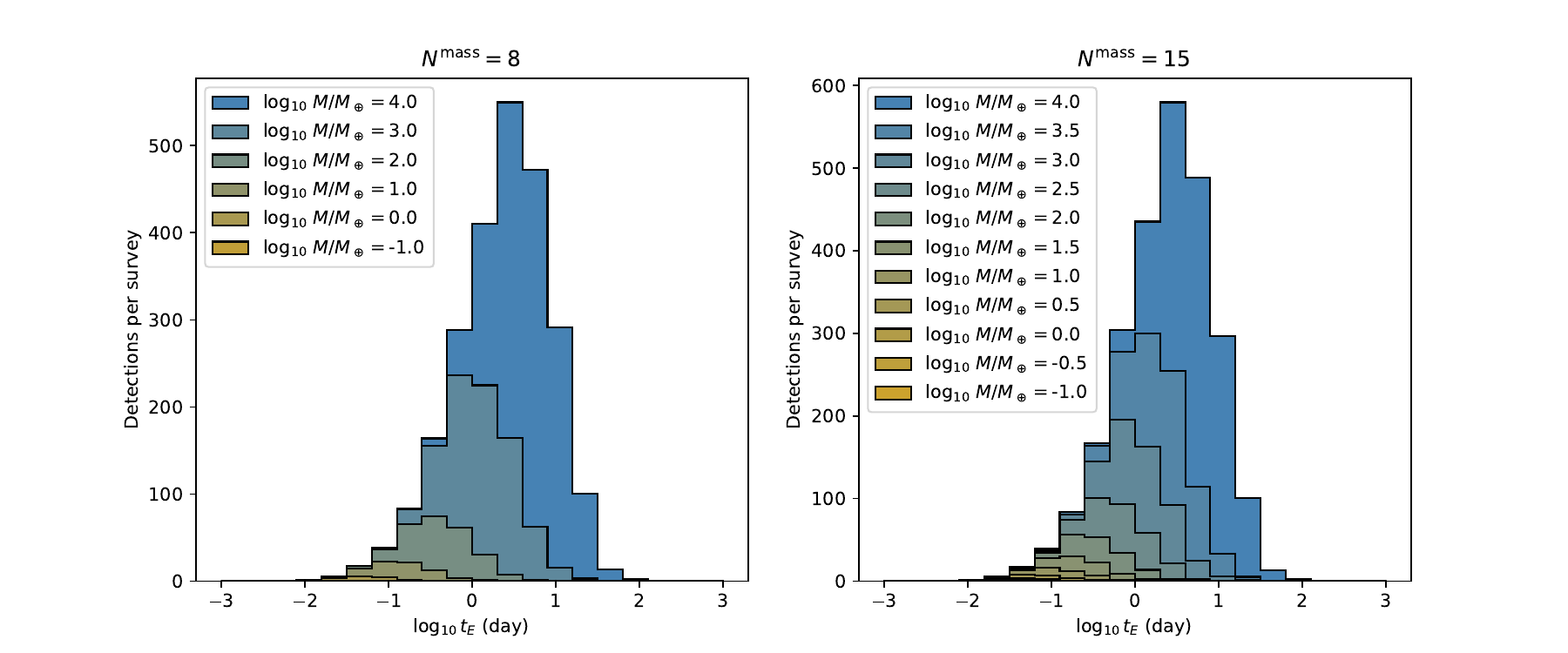}
\caption{Same as Fig. \ref{fig:tEhists} but for a log-uniform mass function normalized to 1 FFP per star per dex and $N^{\text{mass}} = 8$ (left) and $15$ (right). In the righthand plot, most of the bins share roughly equal contributions of nearby masses, leading to bin confusion (see Sec. \ref{sec:binconfusion})
\label{fig:5v15hist}}
\end{figure}

\begin{figure}[ht!]
\plotone{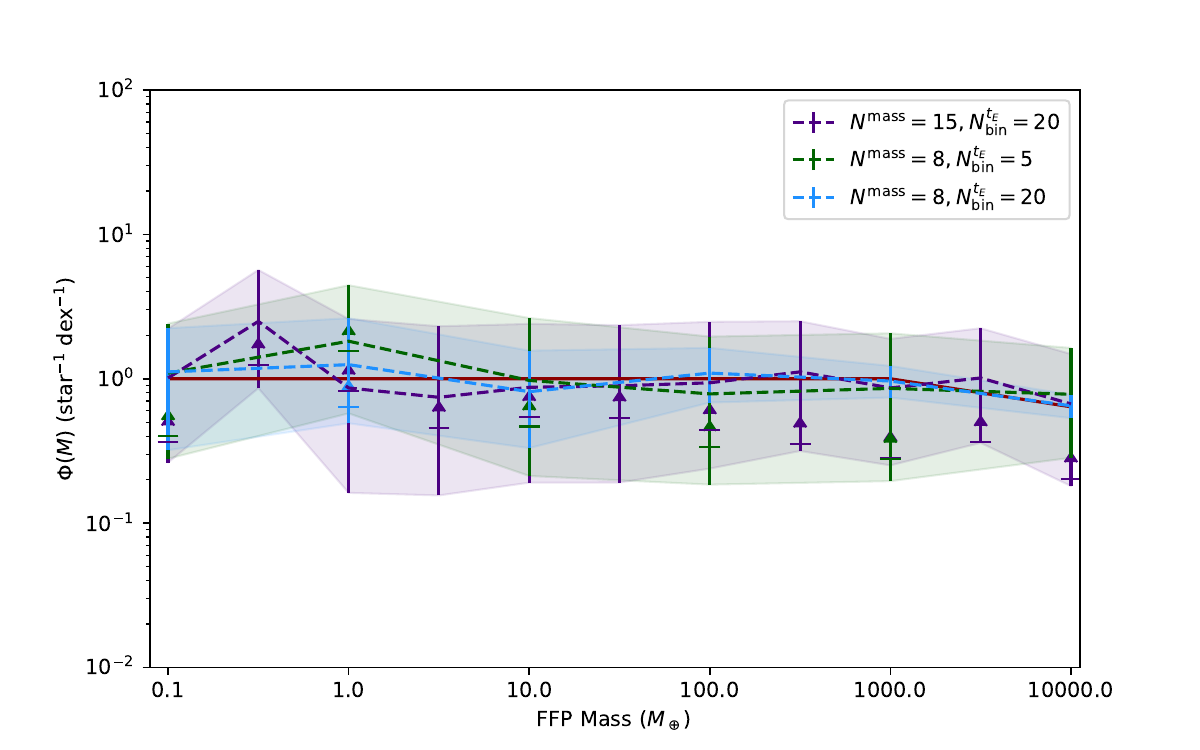}
\caption{Reconstructions of log-uniform mass function for various hyperparameters. Our fiducial values are shown as a blue band. Increasing the number of mass values sampled leads to bin confusion and larger uncertainties (purple). The same is true for decreasing the number of $t_E$ bins (green). Arrows indicate the threshold below which the reconstruction is dominantly prior-supported, hence statistically indistinguishable from zero. The red line is the truth mass function.
\label{fig:binconfusion}}
\end{figure}

If, however, the number of mass values is significantly increased, any given $t_E$ bin contains a roughly even distribution of events from adjacent mass values, reducing the sensitivity (while increasing the mass resolution of the reconstruction). This is demonstrated in in Fig. \ref{fig:5v15hist}, which shows equivalent stacked histograms for the reconstruction of log-uniform FFP mass function, normalized to 1 FFP per star per dex. In the reconstruction in the left of Fig. \ref{fig:5v15hist}, we have adopted $N^{\text{mass}} = 8$ mass values, while in the right, the reconstruction has been parameterized with $N^{\text{mass}} = 15$ mass values. These plots show that by increasing the number of mass values, each individual $t_E$ bin has a more even mixture of different masses, hence the variation of any given $\Phi(M_k)$ can be compensated by a change in $\Phi$ for nearby mass values. As a result, the uncertainty increases, as can be seen in Fig. \ref{fig:binconfusion}, where both reconstructions are plotted on top of one another. The purple band, corresponding to $N^{\text{mass}} = 15$, has significantly larger uncertainties than the blue band, corresponding to $N^{\text{mass}} = 8$. (We additionally include a reconstruction for $N^{\text{mass}} = 8$ and $N^{t_E}_{\text{bins}} = 5$ in green on Fig. \ref{fig:binconfusion} to demonstrate that a similar behavior occurs for a low number of $t_E$ bins.) Hence there is a trade-off: narrower features in the mass function could in principle be uncovered for a larger value of $N^{\text{mass}}$, however the uncertainty increases as a result of these adjacent-mass degeneracies. The optimal balance of these effects depends on the (unknown) underlying mass function, hence cannot be optimized in advance. However, given Roman's ultimate set of FFP detections, these hyperparameters can be varied and different reconstructions performed to find a good balance of resolution and uncertainty.

\subsection{The inclusion of other parameters}
\label{sec:rho}

Though the above results demonstrate that the distribution of $t_E$ alone is potentially able to distinguish various models of the FFP mass function, $t_E$ is not the only observable which Roman may measure. For low-mass lenses, Roman will also be sensitive to $\rho$, the finite-source parameter. It is therefore natural to ask whether using the joint $t_E - \rho$ distribution provides a better handle with which to reconstruct the mass function.

\begin{figure}[ht!]
\plotone{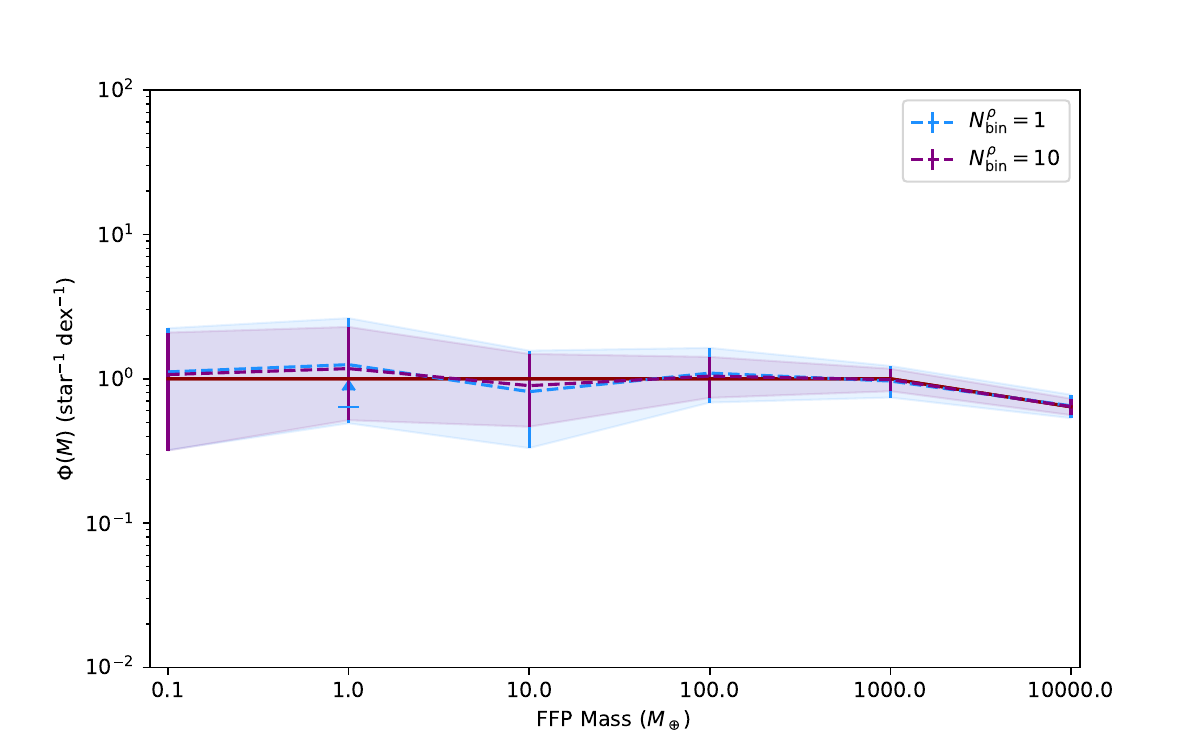}
\caption{Reconstructions of log-uniform mass function both with (purple) and without (blue) $\rho$ information included. Here, $N^{t_E}_{\text{bin}} = 20$ in both cases and $N^{\rho}_{\text{bins}} = 10$ for the fit to the joint $t_E - \rho$ distribution (purple). The improvement is only marginal due to the similar discriminatory power of these two parameters. (See Sec. \ref{sec:rho}.)
\label{fig:rhoplot}}
\end{figure}

\begin{figure}[ht!]
\plotone{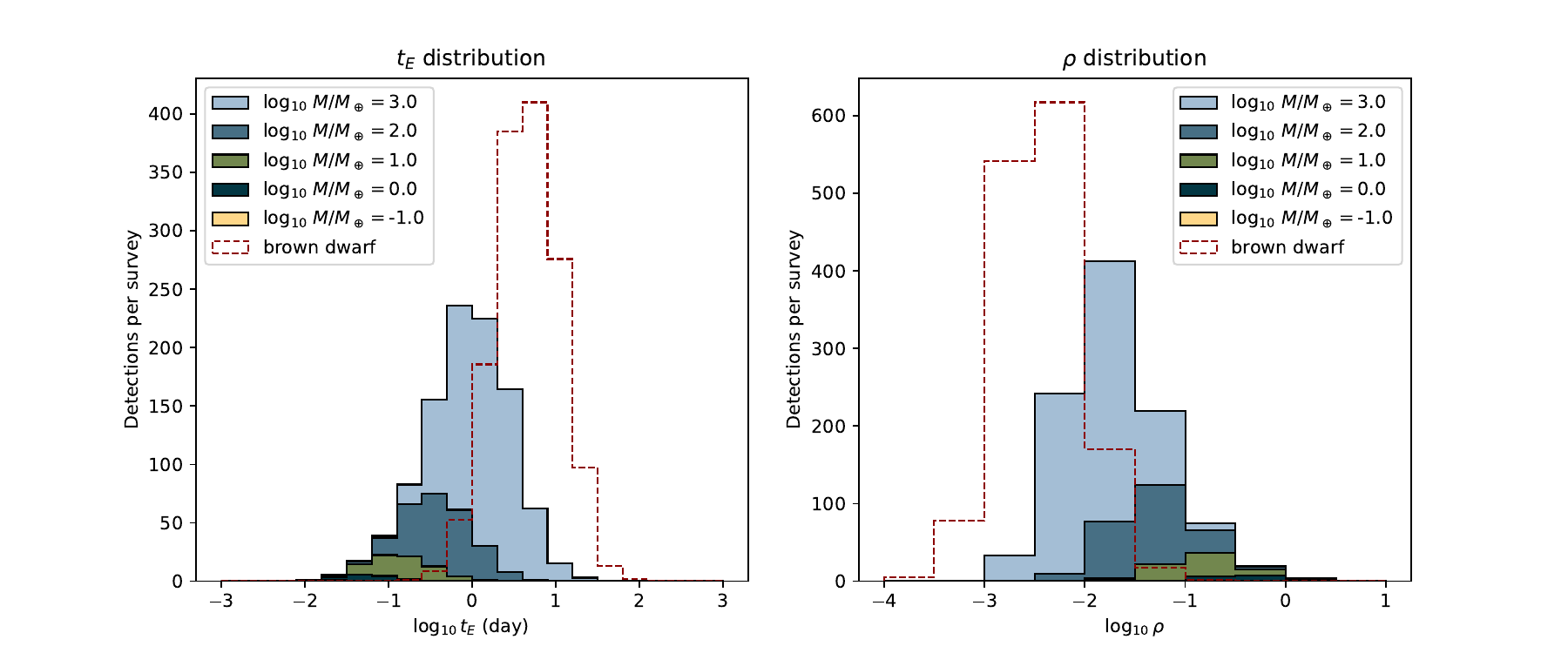}
\caption{Stacked histogram of detected events showing the respective contribution to each $t_E$ (left plot) and $\rho$ (right plot) bin of FFPs of different masses assuming a log-uniform FFP mass function normalized to 1 per star per dex. Here, $N^{\text{mass}} = 8$, $N_{\text{bins}}^{t_E} = 20$, and $N_{\text{bins}}^{\rho} = 10$. The relative distribution of events is similar between the two parameters, hence there is not an appreciable improvement in sensitivity when $\rho$ information is included in the fit.
\label{fig:logunihist}}
\end{figure}

The answer is that while the inclusion of $\rho$ does improve the uncertainties on the reconstruction, the improvement is marginal. See, for example, Fig. \ref{fig:logunihist}, which shows a comparison between a reconstruction of a log-uniform 1 per star per dex mass function using solely $t_E$ information (blue band) and using both $t_E$ and $\rho$ information (purple band). Here, $N^{t_E}_{\text{bin}} = 20$ in both cases and $N^{\rho}_{\text{bins}} = 10$ in the latter and we have extended the method described in Sec. \ref{sec:methods} from a one-dimensional $t_E$ histograms to a two-dimensional $t_E - \rho$ histogram, again evenly sampled in log-space and running from $\log_{10} (t_E/\text{day}) \in [-3, 3], ~\log_{10}(\rho) \in [-3, 3]$. 
We see that while the reconstruction including $\rho$ has slightly smaller uncertainties, the improvement is not large. This is due to the fact that most of the information provided by the inclusion of the $\rho$ distribution is already contained within the $t_E$ distribution. Once again, we can see this through the stacked histogram plots presented above, this time for both the $t_E$ and $\rho$ (Fig. \ref{fig:logunihist}) distributions of a log-uniform mass function. These plots show that there are no masses for which the $\rho$ distribution provides significantly greater discriminatory power in comparison to the $t_E$ distribution, or, put another way, there is no particular $\rho$ bin that is strongly dominated by a mass which does not dominate in a $t_E$ bin as well. Contamination by brown dwarfs does not appreciably change this conclusion, as brown dwarfs typically contaminate the same high-mass range in both $t_E$ and $\rho$ (see Fig. \ref{fig:logunihist}), hence the addition of $\rho$ does not provide significantly more information by which to constrain the mass function.

Additionally, note that unlike $t_E$, which is expected to be largely well-measured in Roman light curves (i.e. $\delta t_{E,i} < $ $t_E$ bin width; see Fig. \ref{fig:deltatE}), $\rho$ is subject to considerable uncertainty and is often undetectable. This induces a large associated uncertainty on the bin heights of the $\rho$ distribution, resulting in a more uncertain reconstruction. Due to these factors, the inclusion of $\rho$ provides only a small improvement on the \textit{statistical} uncertainty of the mass function reconstruction over using $t_E$ distributions alone. In practice, however, the inclusion of $\rho$ may provide a significant enhancement, as it may reduce \textit{systematic} uncertainty on a particular Galactic model and enable the mitigation against potential false positives. These cases will be explored in future work.

\subsection{Systematic uncertainties}
\label{sec:syst}

So far, our results have shown confidence intervals based \textit{purely on statistical uncertainties}, both from sampling variance in binned histograms and from the imprecise measurement of $t_E$ and $\rho$ from a given light curve. However, as mentioned briefly in Sec. \ref{sec:sims}, systematic uncertainties may also arise from, e.g., the adoption of a particular Galactic model. It is an implicit assumption of our methodology that the Galactic model employed is a perfect model of the Galaxy; this assumption introduces an associated systematic uncertainty. We explore this uncertainty by performing an identical analysis to that presented above however with variations on the underlying Galactic model. 

Recall that our framework assumes the Galactic model presented in \citet{SynthPop} (hereafter, SynthPop) to be a perfect model of the Galaxy. As such, the construction of the $n_j^{t_E}(\Phi(M_k))$ in Eq. \ref{eq:likelihood} is based upon the mapping between mass bins and $t_E$ distributions associated with this Galactic model. In all of the above tests, the observed data was \textit{also} generated from this model, i.e., both the histograms $n_j^{t_E, \text{obs}}$ and $n_j^{t_E}(\Phi(M_k))$ were generated from SynthPop. In this section, we will instead modify this Galactic model in simple analytic ways to generate the observed data $n_j^{t_E, \text{obs}}$ while still basing our inferred distribution $n_j^{t_E}(\Phi(M_k))$ on the unchanged SynthPop. This model mismatch is intended to capture the effects of having an imperfect knowledge of the true Galactic model.

We perform two variations on SynthPop. The first is to modify the number of lenses in the Galaxy. We introduce a distance-dependent multiplicative factor to the lensing rate of the form
\begin{equation}
\label{eq:changeN}
    10-9\times\left(\frac{|8\,\text{kpc} - D_L|}{8\,\text{kpc}}\right)
\end{equation}
which is 1 at Earth ($D_L = 0$ kpc) and 10 at the Galactic Center ($D_L = 8$ kpc). This represents a dramatic increase in the number of lenses towards the Galactic Center, well beyond realistic values for the mismodeling in this abundance. However, we choose to adopt this in order to demonstrate that the reconstruction is fairly robust even under such extreme variations.

The second modification is to introduce a distance-dependent change to the relative proper motion of lenses. For this, we introduce a multiplicative factor to both the lensing rate \textit{and} the values of $t_E$ (as a change in proper motion also causes a change in $t_E$) of the form 
\begin{equation}
\label{eq:changemurel}
    0.8+0.2\times\left(\frac{|8\,\text{kpc} - D_L|}{8\,\text{kpc}}\right)
\end{equation}
which is 1 at Earth and 0.8 at the Galactic Center. Again, this is likely far larger than realistic modeling uncertainties, but serves as an example.

\begin{figure}[ht!]
\plotone{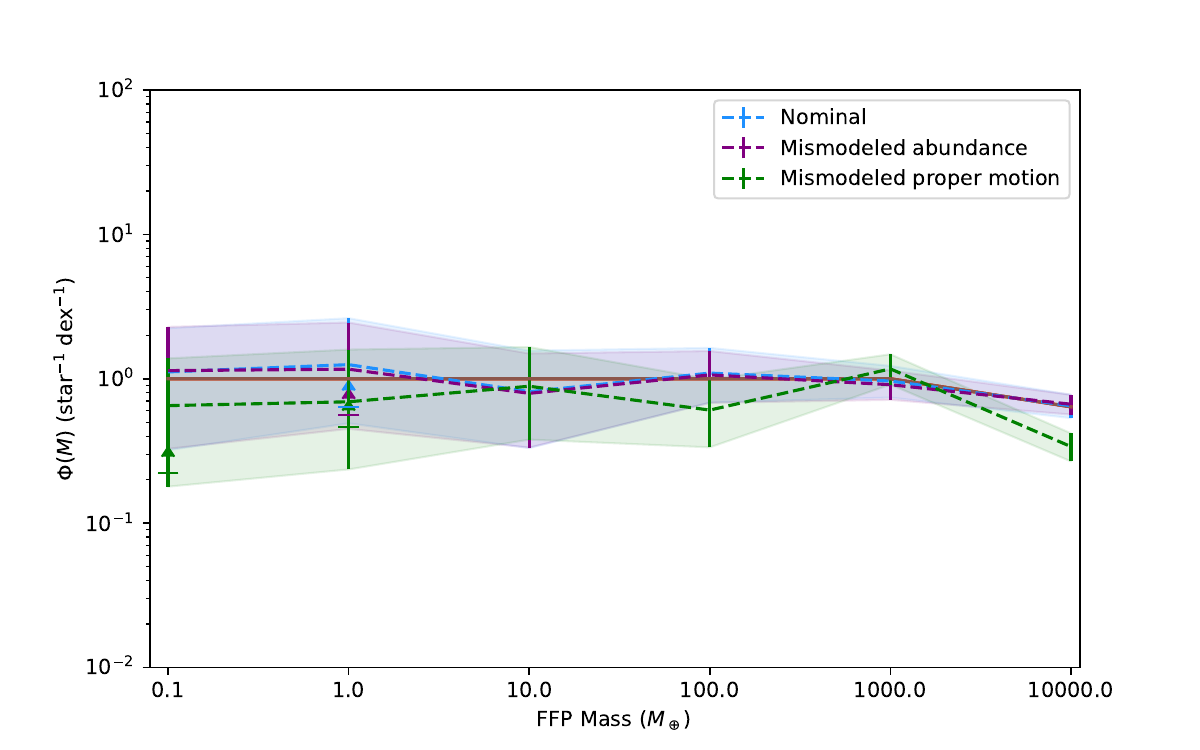}
\caption{Reconstructions of log-uniform mass function for three sets of observed data generated using different Galactic models. In blue, both the inferred and observed distribution were generated from the Galactic model of \citet{SynthPop} (SynthPop). In purple (green), the inference is still based upon SynthPop while the observed distribution was generated according to Eq. \ref{eq:changeN} (Eq. \ref{eq:changemurel}). This exercise demonstrates the systematic errors that can arise from an imperfect knowledge of the true Galactic model.  (See Sec. \ref{sec:syst}.)  
\label{fig:logunimismatch}}
\end{figure}

The primary result is shown in Fig. \ref{fig:logunimismatch}. We show in blue the reconstructed mass function in the case where both the inference distribution and observed distribution were generated from the nominal SynthPop model. In purple and green, we show the cases in which the inference is still based upon nominal SynthPop while the observed distribution was generated with the distance-dependent modifications given by Eqs. \ref{eq:changeN} and \ref{eq:changemurel} respectively, with a log-uniform 1 per star per dex FFP mass function for all cases. All of these reconstructions were performed with $N^{\text{mass}} = 8$ and $N^{t_E}_{\text{bins}} = 20$. 

The results of this exercise show that Galactic mismodeling will introduce systematic uncertainties into the analysis (e.g. the underestimated mass function at 100 and $10^4 \,M_{\oplus}$ in the varied $\mu_{\text{rel}}$ case), though for the majority of masses, the statistical uncertainties arising from low yields will likely dominate most realistic modeling uncertainties. However, if overall FFP abundance is much higher than that predicted by the log-uniform mass distribution, then systematic uncertainties will begin to play a role across a wider range of masses.

The results of this section motivate joint precursor and contemporaneous observations of Roman's field of view that provide key complementary insight into the FFP  mass function. Precursor observations will help refine our present models of the Galaxy towards the Bulge, reducing systematic uncertainties, and contemporaneous observations may provide parallax measurements of FFPs, allowing the mass of such events to be measured directly. Both of these will prove critical in successfully reconstructing the FFP mass function from Roman's observations in the face of low statistics and model-induced systematic bias.

\section{Conclusions and Future Work} \label{sec:conc}

In this paper, we have explored the potential for the upcoming Nancy Grace Roman Space Telescope to reconstruct the mass distribution of free-floating planets in the Galaxy. We find that by leveraging population-level statistics, Roman will have the potential to discriminate between competing models of the FFP mass function in phenomenologically interesting mass regimes. In particular, if the true underlying mass function corresponds to the power-law fit to existing ground-based observations presented in \citet{Sumi2023}, then Roman has the potential to significantly improve upon the existing measurement of the abundance at sub-Earth masses. However, if the mass function instead follows the planet population synthesis-motivated form presented in \citet{Coleman_2025}, Roman will likely require joint observations in order to to discern key non-monotonic features, such as a peak near $8\,M_{\oplus}$, that would provide new insight into the dominant processes that eject free-floating planets and the environments in which they form.

This work opens many compelling avenues for future research. Though we have explored Roman's ability to reconstruct the mass function with $t_E$ and $\rho$, the use of other additional observables may also significantly improve upon the results. If, for example, contemporaneous observations of Roman's microlensing events are performed, either by ground- or space-based missions, the measurement of microlensing parallax could significantly constrain the underlying mass of lenses \citep{Bachelet2019,Ban2020,Bachelet2022,Ban2023}, providing an anchor for the population-level reconstruction algorithm. Such direct mass measurements provide an important complementary probe of the mass function that does not rely on the assumption of a perfect Galactic model. By leveraging such observations along with our population-level framework, systematic uncertainties in the reconstruction of the mass function can be significantly reduced.

Additionally, the analysis presented in this paper has focused on the case where there is perfect discrimination of FFP microlensing events from potential backgrounds. This may no longer be a robust assumption for space-based missions, hence a more complete analysis would marginalize over this potential contamination as well. At present, the potential false positives for short-duration microlensing are not well understood \citep{Kunimoto2024}. However, precursor observations may significantly improve our understanding of such signals, providing an opportunity to extend our reconstruction framework to incorporate them.

Finally, these results are more broadly applicable to the search for isolated non-luminous bodies other than free-floating planets. Primordial black holes \citep{Carr_2022}, for example, are a compelling dark matter candidate for which the methodology presented in this paper would be able to reconstruct the mass function. This requires a non-trivial conversion between the FFP distribution and dark matter distribution \citep{Derocco_2023} and the marginalization over the free-floating planet background \citep{DeRocco_2024}. A complete exploration of the detection prospects is left to future work.

In conclusion, the Nancy Grace Roman Space Telescope will open up a new era of discovery not only for bound exoplanet demographics, but unbound planets as well, and will provide our first glimpse into the origins of these mysterious rogue worlds.

\begin{acknowledgments}
\textit{Acknowledgements:} Much of this work was supported by the dedicated effort of Scott Perkins, whose contributions to ensuring the robustness of the statistical methodology were invaluable. WD further wishes to thank Greg Olmschenk and David Bennett for useful comments on the framework and relation to existing studies. WD was supported by NSF grant PHY-2210361 and the Maryland Center for Fundamental Physics. MTP was supported by NASA awards 80NSSC24M0022 and 80NSSC24K0881, and FZ by the former. Work by SAJ was supported by NASA Grant 80NSSC24M0022 and by an appointment to the NASA Postdoctoral Program at the NASA Jet Propulsion Laboratory, administered by Oak Ridge Associated Universities under contract with NASA. This work was performed under the auspices of the U.S. Department of Energy by Lawrence Livermore National Laboratory under Contract DE-AC52-07NA27344. PM was supported by the LLNL-LDRD Program under Project 22-ERD-037. The document release number is LLNL-JRNL-2003768. Computational work was carried out at the Advanced Research Computing at Hopkins (ARCH) core facility  (rockfish.jhu.edu), which is supported by the National Science Foundation (NSF) grant number OAC1920103.
\end{acknowledgments}

%

\vspace{5mm}
\facilities{Advanced Research Computing at Hopkins (ARCH)}


\software{NumPy \citep{numpy}, Pandas \citep{pandas}, SciPy \citep{scipy}, GULLS \citep{Penny2019}
          }



\appendix

\section{Glossary of symbols} \label{app:glossary}

In Table \ref{tab:symbols}, we include a glossary of the notation used in Sec. \ref{sec:methods}.

\begin{table}[h]
    \centering
    \caption{Definition of Symbols Used in the Analysis}
    \begin{tabular}{c l}
        \hline
        Symbol & Definition \\
        \hline
        $N_{\text{sim}}$ & Total number of simulated events \\
        $w_i$ & GULLS weight assigned to event $i$ in simulation \\
        $N_{\text{pred}}$ & Predicted number of FFP microlensing events per survey \\
        $\Phi(M)$ & FFP mass function \\
        $\tilde{w}_i$ & Rescaled event weight for non-uniform mass function \\
        $N^{\text{det}}_{\text{sim}}$ & Number of simulated events after applying detection cuts \\
        $N^{\text{det}}_{\text{pred}}$ & Predicted number of detectable microlensing events per survey \\
        $N_{3\sigma}$ & Number of $3\sigma$-significant observations in an event \\
        $\Delta \chi^2$ & Difference in goodness-of-fit between baseline and Paczyński models \\
        $N^{t_E}_{\text{bins}}$ & Number of bins in $t_E$ histogram \\
        $N^{\rho}_{\text{bins}}$ & Number of bins in $\rho$ histogram \\
        $N^{\text{mass}}$ & Number of mass values in parameterized mass function\\
        $n^{t_E}_j$ & Expected number of events in $t_E$ bin $j$ \\
        $p_{i,j}$ & Probability of event $i$ being measured in bin $j$ \\
        $\sigma_j^0$ & Poisson uncertainty in bin $j$ \\
        $\sigma_j^{\delta t_E}$ & Uncertainty incorporating $t_E$ measurement errors \\
        $\sigma_j$ & Larger (relative) uncertainty of $\sigma_j^0$ and $\sigma_j^{\delta t_E}$\\
        \hline
    \end{tabular}
    \label{tab:symbols}
\end{table}

\bibliography{sample631}{}
\bibliographystyle{aasjournal}



\end{document}